\documentclass[12pt]{iopart}
\usepackage[utf8]{inputenc}
\usepackage{graphicx}
\usepackage{textcomp}
\usepackage{placeins} 	
\begin{document}
\title{Nonlinear conductivity in CaRuO$_3$ thin films measured by short current pulses}

\author{Sven Esser$^1$, Sebastian Esser$^1$, Christian Stingl$^1$ and Philipp Gegenwart$^1$}

\address{$^1$I. Physikalisches Institut, Georg-August-Universit\"at G\"ottingen, 37077 G\"ottingen, Germany}

\ead{pgegenw@gwdg.de}

\begin{abstract}
Metals near quantum critical points have been predicted to display universal out-of equilibrium behavior in the steady current-carrying state.
We have studied the non-linear conductivity of high-quality CaRuO$_3$ thin films with residual resistivity ratio up to 57 using  micro-second short, high-field current pulses at low temperatures.
Even for the shortest pulses of 5\,\textmu s, Joule heating persists, making it impossible to observe a possible universal non-linearity.
Much shorter pulses are needed for the investigation of universal non-linear conductivity.
\end{abstract}


\section{Introduction}

A quantum critical point~(QCP) arises when the critical temperature of a phase transition is suppressed to $T=0$ by means of an external control parameter such as pressure, magnetic field or doping~\cite{sondhi}.
In contrast to a phase transition at finite temperature, which is driven by thermal fluctuations, a quantum phase transition (QPT) is characterized by quantum fluctuations between the ordered and disordered state.
Although the QCP at $T=0$ is experimentally not accessible, these quantum fluctuations drastically influence many physical properties even at finite temperatures.
Examples include non-Fermi-liquid (NFL) behavior in resistivity or electronic specific heat~\cite{gegenwart_naturephysics_2008} and divergences in the thermal~\cite{kuechler_ceni2ge2_grueneisen_PRL_2003} and magnetic~\cite{tokiwa_prl_2009} Gr\"uneisen ratios.
Quantum critical materials display thermodynamic equilibrium fluctuations completely controlled by temperature and a non-thermal parameter, leading to universal energy over temperature scaling in wide regions of phase space.
Importantly, it has been suggested that universality may extend also to out-of-equilibrium dynamics in driven quantum critical states beyond linear response~\cite{dalidovich2004nonlinear}.
In particular it has been proposed that transport properties of quantum critical metals can be understood using an effective temperature induced by the current and non-linear electrical resistivity is determined by universal exponents of the underlying QCP~\cite{hogan}.
Experimental observation of universal non-linear non-equilibrium behavior far from thermal equilibrium would be also of general importance for  understanding of energy conversion processes.

Hogan and Green~\cite{hogan} have considered the steady-state nonlinear conductivity of a quantum critical metal at large electric fields.
It is expected when the energy gained by an electron from the electric field between scattering events exceeds the temperature.
Above a critical field $E_1\approx T_0/l_\mathrm{tr}$, with the temperature~$T_0$ and the electronic transport mean free path~$l_\mathrm{tr}$, they found a universal relation between current density $j$ and electric field $E$, given by
\[
j\propto E^{(z-1)/[z(1+\alpha)-1]}~.
\]
Here $z$ denotes the dynamical critical exponent which equals 3 for ferromagnetic and 2 for antiferromagnetic quantum critical metals and the parameter $\alpha$ is the temperature exponent of the electrical resistivity in thermal equilibrium, $\rho(T)=\rho_0 + AT^\alpha$, experimentally often found between 1 and 1.5.

In order to minimize the critical field $E_1$, low temperatures and clean samples, which are indicated by high residual resistivity ratio (RRR), are needed.
In addition, a temperature rise by Joule heating increases the sample resistance and therefore trivially leads to a nonlinear $j$-$E$-curve.
This requires that the sample must be very well thermally coupled in order to transport the dissipated heat.
We are not aware of any experimental studies on possible non-linear conductivity behavior in quantum critical metals.
The need of high electric fields and a very good thermal coupling to the temperature bath requires the usage of thin film samples for which high surface-to-volume ratio ensures an efficient heat transport.
Furthermore, a well-ordered stoichiometric material is need for the requirement of a large mean free path.
There are only very few clean metals near QCPs and even much less can be prepared as thin films.

We attempt to meet these stringent requirements by choosing for our experiments thin films of the perovskite ruthenate CaRuO$_3$.
Studies on the series Sr$_{1-x}$Ca$_x$RuO$_3$ suggest that CaRuO$_3$ is close to a ferromagnetic QCP \cite{schneider, cao}.
We have recently shown, that  CaRuO$_3$ displays a very fragile Fermi liquid ground state and extended temperature range in which the electrical resistivity follows a $\rho(T)=\rho_0+AT^{3/2}$ behavior~\cite{schneider2013low}.
Most importantly, very high-quality epitaxial unstrained thin films with resistivity ratio above 50, allowing the study of quantum oscillations, can be prepared by metal-organic aerosol deposition~\cite{schneider2013low}.
In order to reduce Joule heating we perform resistivity measurements with short pulses in the microsecond range.

Our paper is outlined as follows.
In section \ref{sec:experimental} we will discuss sample preparation and characterization, examine the heat flows in our setup and describe the circuit used for producing the short current pulses.
We show that the pulse shapes can be well analysed in order to extract the sample resistance.
Then in section \ref{sec:results} we show the experimental results for the pulse length dependence of the electrical resistivity and the shape of the measured $j$-$E$-curve.
Finally we discuss our results and establish a relationship to the theoretical predictions made by Hogan and Green.


\section{Experimental Details}\label{sec:experimental}

\begin{figure}
	\centering
	\includegraphics[width=0.5\textwidth]{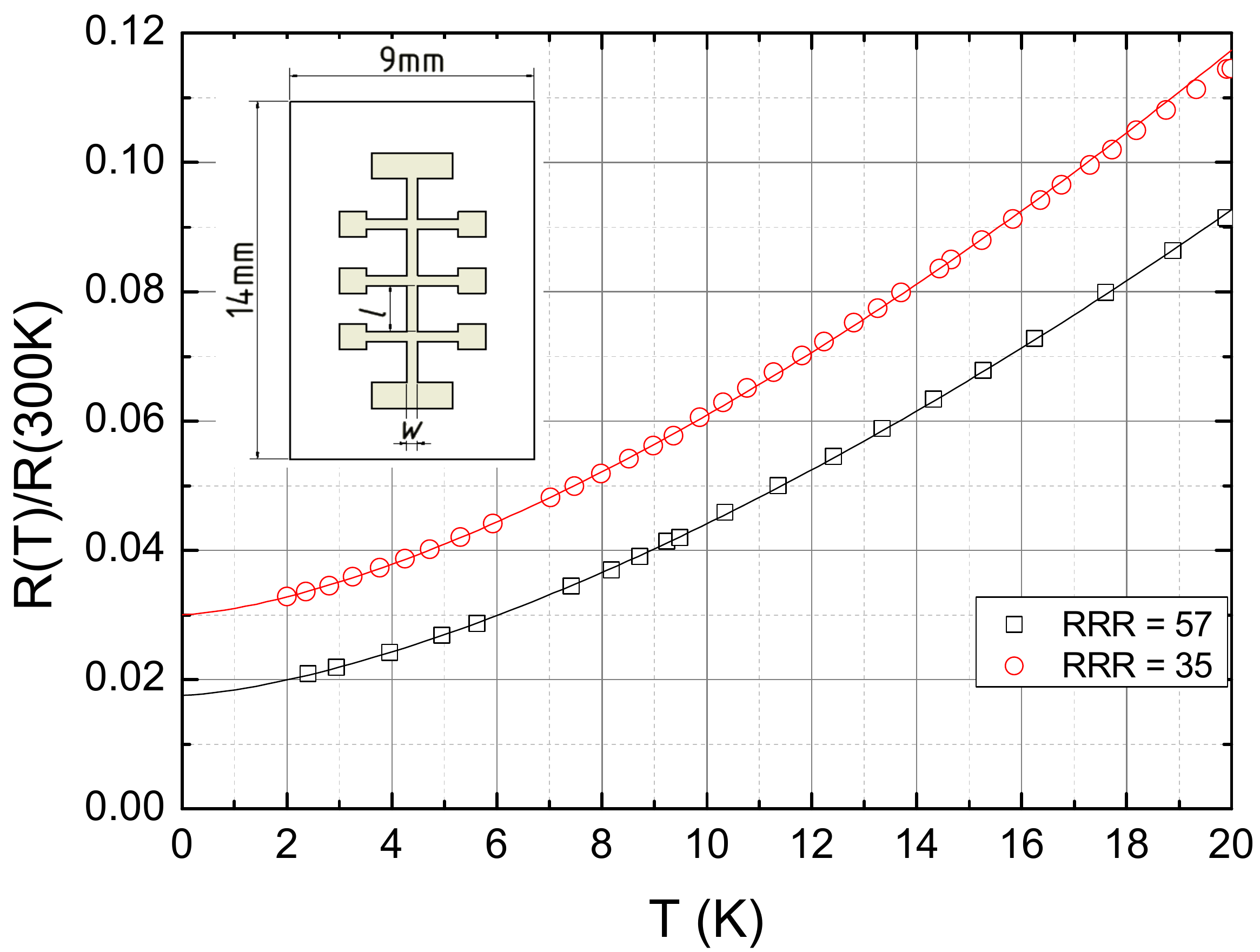}
	\caption{Resistivity as a function of temperature for CaRuO$_3$ thin films with RRR~$=57$ and 80\,nm thickness (black squares) and RRR~$=35$ and 43\,nm thickness (red circles) below 20\,K.
	The lines display~$T^{3/2}$ behavior and the inset shows the geometry of a structured sample with length $l$ and width w of the measurement path.}
	\label{fig:sample_quality}
\end{figure}

\begin{table}
\label{tab:sample_data}
\caption{Geometric parameters length $l$ and width w of the measurement path and thickness $d$ of the thin film as well as the calculated residual resistivity ratio (RRR) and electronic transport mean free path~$l_\mathrm{tr}$ for both CaRuO$_3$ samples.}
\begin{center}
\begin{tabular}{@{}lllll}
\br
RRR&$l$ (mm)&w (\textmu m)&$d$ (nm)&$l_\mathrm{tr}$ (nm)\\
\mr
57&2&90&80&21.24\\
35&1.5&408&43&13.04\\
\br
\end{tabular}
\end{center}
\end{table}

The CaRuO$_3$ thin films are grown on vicinal NdGaO$_3$ substrates by the metal-organic aerosol deposition (MAD) technique, which is described in \cite{MAD, moshnyaga}, and structured by argon ion etching.
X-ray diffraction measurements prove the phase purity of our samples.
For the subsequent investigation of the sample quality we performed resistivity measurements using a Physical Property Measurement System (PPMS) (see Fig.~\ref{fig:sample_quality}).
Both studied samples display a $T^{3/2}$ dependence, similar as found in~\cite{schneider2013low} for temperatures below 20 K, indicating NFL behavior in the electrical resistivity.

The pulsed current measurements were performed in a $^4$He evaporation cryostat.
The samples (substrate and thin film) were mounted on a copper block located directly in the liquid helium bath.
By pumping the bath, the temperature could be reduced from 4.2\,K to $\approx 1.4$\,K.

\begin{figure}
	\centering
	\includegraphics[]{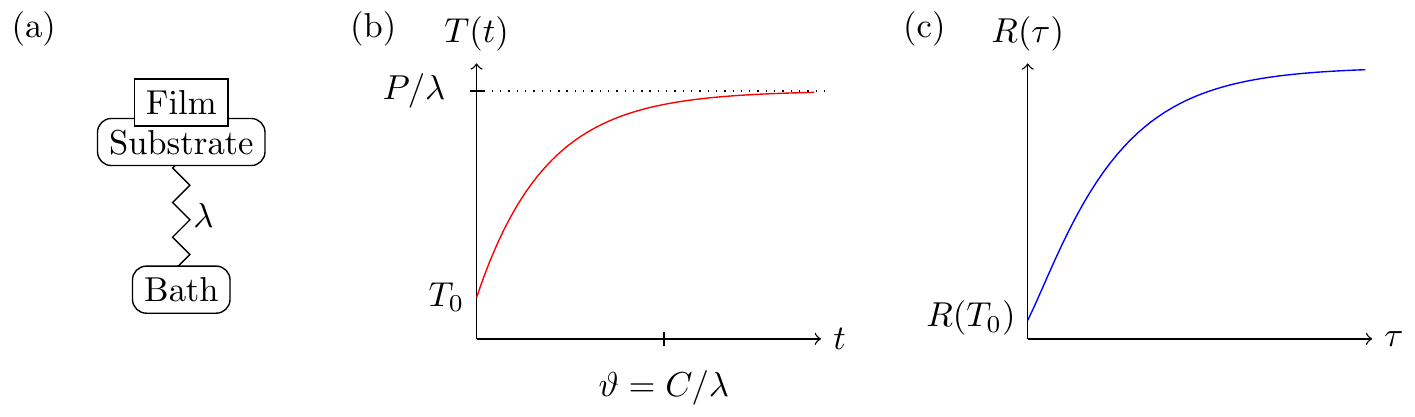}
	\caption{(a) Schematic diagram of the expected thermal transport.
	The substrate acts as a heat sink for the thin film.
	Heat is transferred to the $^4$He bath through a thermal conductivity~$\lambda$.
	(b) The film temperature increases as a function of time due to Joule heating.
	(c) Accordingly, the measured resistance will increase with pulse length~$\tau$.}
	\label{fig:resistivity_schematic}
\end{figure}

For the thermal transport between thin film and surrounding $^4$He bath, we make the optimistic assumption that the epitaxially grown film is well thermally coupled to the substrate, which can therefore act as a heat sink.
Film and substrate then have a combined heat capacity~$C$ which is coupled to the bath by a thermal conductance~$\lambda$ as shown in~Fig.~\ref{fig:resistivity_schematic}a.

When an electric field is applied at~$t=0$ and the bath temperature $T_0$, the sample dissipates the power $P=V\sigma E^2$.
Its temperature will increase as
\[
 \dot{T} = \frac{P}{C}  -\frac{\lambda}{C}\left(T-T_0\right) \quad,
\]
which is solved by
\[
T(t) - T_0 = \frac{P}{\lambda} \left( 1-e^{-t/\vartheta} \right)
\]
with $T(0)=T_0$ and the time constant $\vartheta=C/\lambda$.
For large pulse lengths, a steady state temperature increase $T=P/\lambda$ is reached.
Since resistivity is a function of temperature, it will also increase with time, as illustrated in Fig.~\ref{fig:resistivity_schematic}c.

At low temperatures, the specific heat of NdGaO$_3$ is strongly enhanced by a Schottky contribution from the magnetic Nd$^{3+}$ ground state due to crystal electric field splitting~\cite{schnelle2001}.
At 2\,K, our 0.5\,mm thick substrate has a heat capacity of approximately~15\,\textmu J\,mm$^{-2}$K$^{-1}$.
The Kapitza thermal boundary conductances between a metal and liquid $^4$He (film -- bath) as well as between a metal and an insulator (substrate -- copper block) are of the order of $\lambda \approx 1\,$mW\,mm$^{-2}$K$^{-1}$ at the same temperature~\cite{pobell}.
These values provide a rough estimate of the time constant~$\vartheta\approx 15$\,ms.
In order to minimize Joule heating, we aim to keep the pulse width~$\tau$ small against~$\vartheta$.

\begin{figure}
	\centering
	\includegraphics[width=0.5\textwidth]{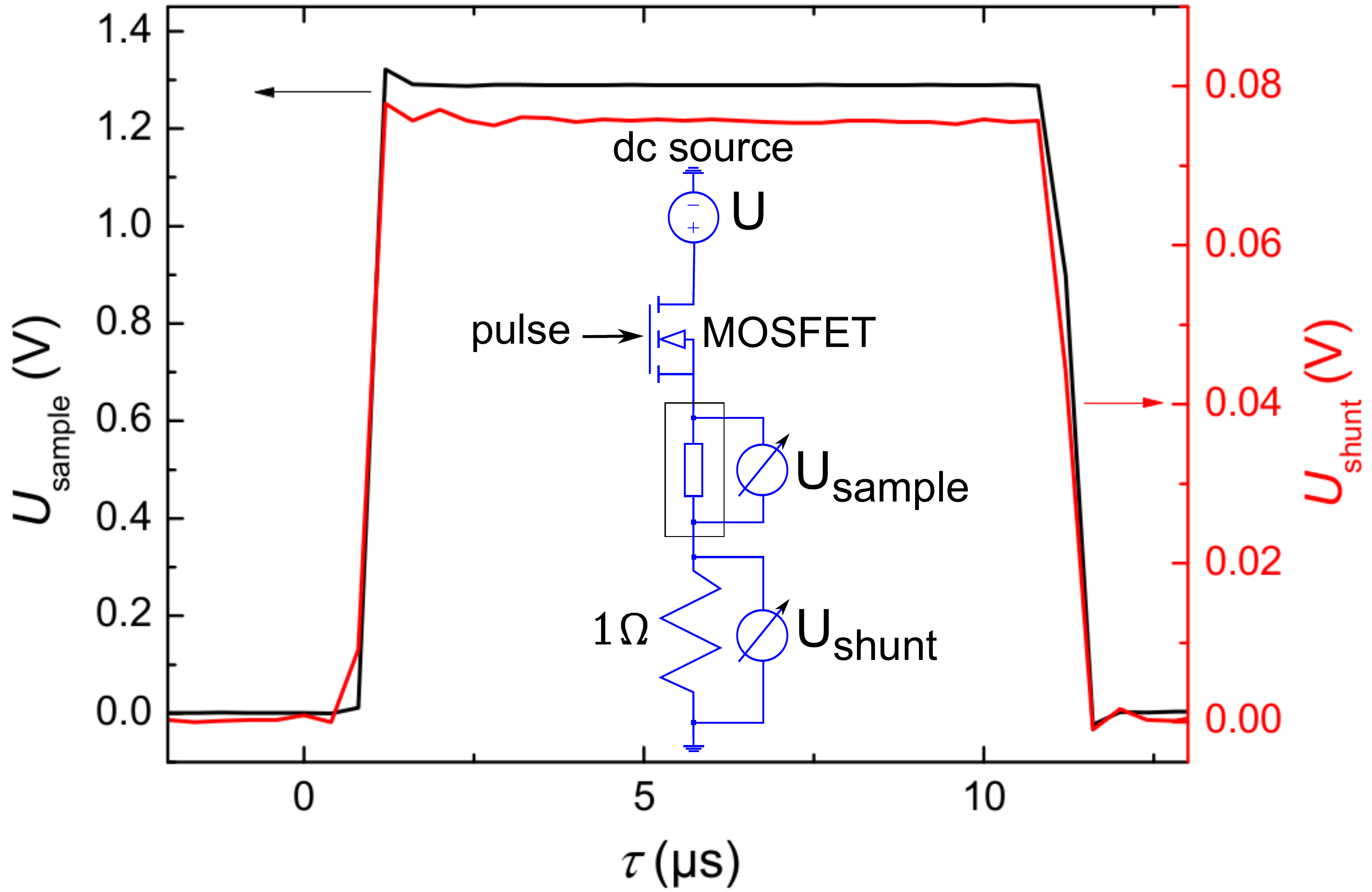}
	\caption{The shape of the single pulse in CaRuO$_3$ sample with RRR~$=57$ (black) and in the shunt (red) at 4.2~K with source voltage $U=15\,$V.
	The inset shows the circuit diagram to realize short current pulses.
	The black rectangle shows the part located inside the cryostat.}
	\label{fig:single_pulse}
\end{figure}

Short current pulses were realized by the circuit shown as inset in Fig.~\ref{fig:single_pulse}.
An adjustable dc voltage source $U$ is connected to the sample via a fast field effect transistor.
The current is measured through the voltage over a high precision $1\,\Omega$~shunt.
A function generator drives the MOSFET gate with pulse length~$\tau$ and a small duty cycle, i.\,e. the delay between consecutive pulses being much larger than~$\tau$.
The voltages $U_\mathrm{sample}$ and $U_\mathrm{shunt}$ are measured differentially with a~14~bit, 2.5~MS/s ADC~card.
Due to the connection in series, the current through the thin film is given by:
\begin{equation*}
I_\mathrm{sample} \equiv I_\mathrm{shunt} = \frac{U_\mathrm{shunt}}{1\Omega}
\end{equation*}
Pulse lengths of $\tau \geq 5\,$\textmu s are possible with the used circuit elements.
A more detailed circuit diagram is shown in the appendix.

Figure~\ref{fig:single_pulse} shows the measured voltages for a pulse length of~$\tau=10\,$\textmu s.
After switching the transistor, the voltage rises quickly to the final value for $U_\mathrm{sample}$  and $U_\mathrm{shunt}$ and reaches a plateau, before dropping quickly to zero as the transistor is turned off.
We analyze the measured waveforms and extract the height of the plateau, accounting for some over- and undershoot at the flanks of the pulse.
By this method, we can determine the voltage and the current for the thin film during a single pulse.


\section{Results}\label{sec:results}

\subsection{Pulse length dependence of the electrical resistivity}

\begin{figure}
	\centering
	\includegraphics[width=0.5\textwidth]{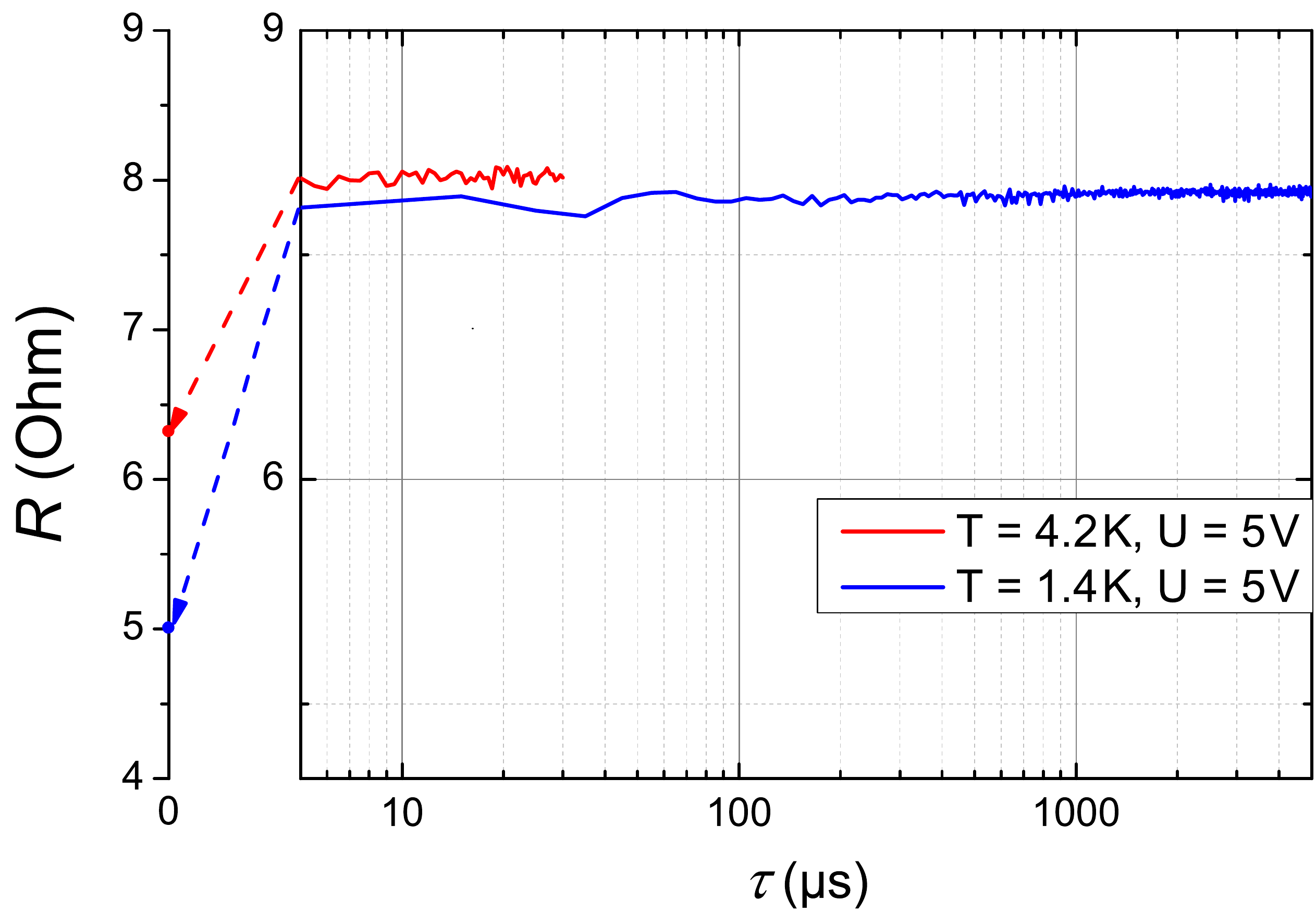}
	\caption{resistance as a function of pulse length~$\tau$ for the CaRuO$_3$ thin film with RRR~$=35$ at $T=4.2$\,K (red) and $T=1.4$\,K (blue).
	The ratio $E/E_1$ is 3/25 and 9/25 at 4.2\,K and 1.4\,K, respectively.
	The left vertical scale displays the resistance from a conventional, low current PPMS measurement, i.\,e. without Joule heating.}
	\label{fig:pulsesweep}
\end{figure}

In order to investigate the pulse length dependence of the electrical resistivity, we vary the pulse length~$\tau$ at a constant dc voltage~$U$ and temperature~$T$.
Figure~\ref{fig:pulsesweep} displays the dependency between resistance and pulse length for the CaRuO$_3$ sample with RRR~$=35$ at 4.2\,K and 1.4\,K, respectively.
At $U=5$\,V, the electric field is still below the criticial field~$E_1$.
The measured resistance is constant for all investigated pulse lengths.
A shape as in Fig.~\ref{fig:resistivity_schematic}c can not be observed.
Arrows indicate the resistivity values determined from a conventional low-current measurement.
As no heating occurs there, these values correspond to a vanishing pulse length~$\tau=0$.
The difference to the expected value is $1.7\,\Omega$ and $2.8\,\Omega$ at $4.2\,$K and $1.4\,$K, respectively.

\begin{figure}
	\centering
	\includegraphics[width=0.5\textwidth]{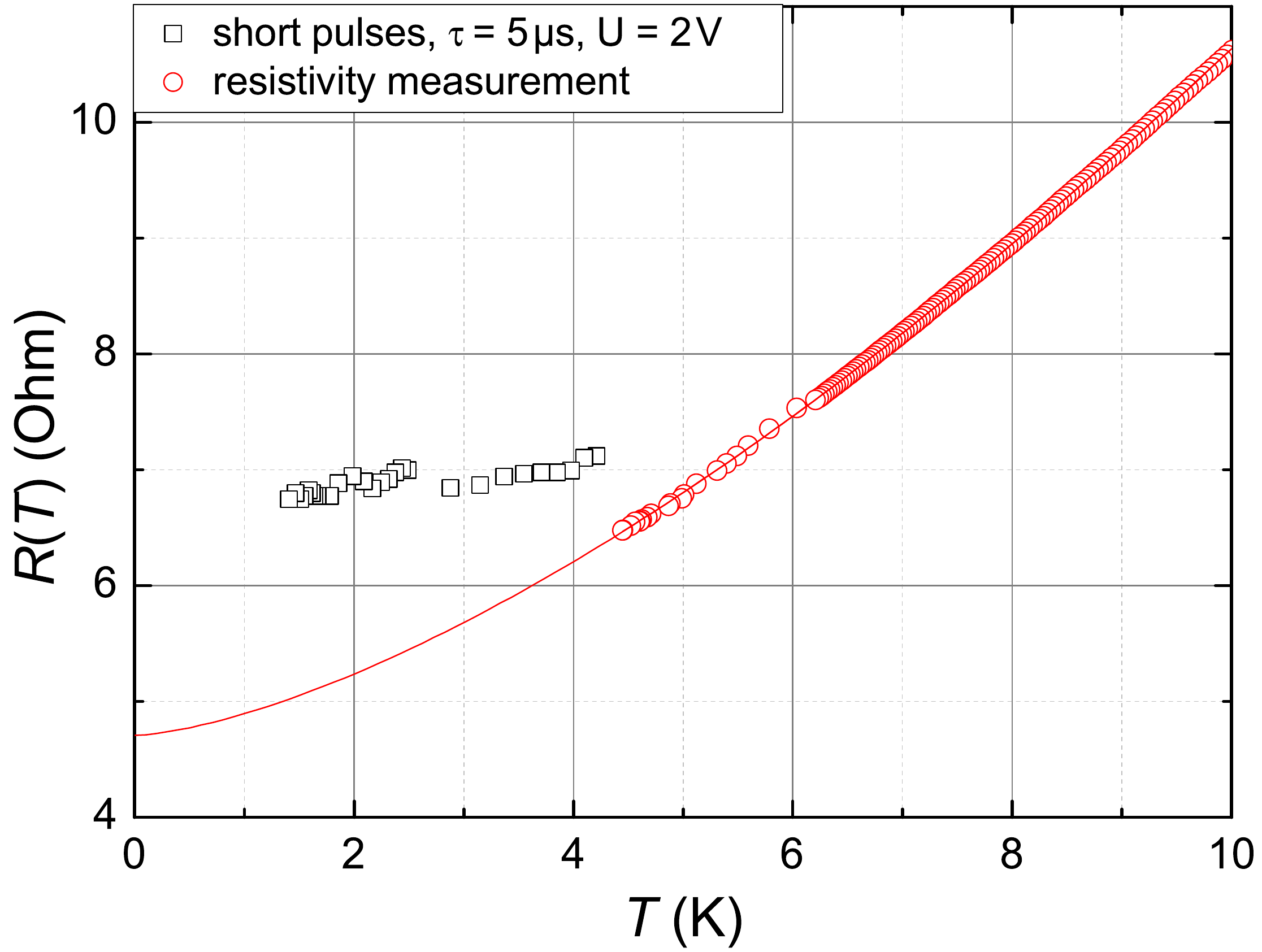}
	\caption{Resistivity vs. temperature of a CaRuO$_3$ thin film with RRR~$=35$ below $10\,$K determined from short pulse measurements at $E=1333\,$V/m and $I=58\,$mA (black squares) and conventional, low-current resistivity measurements at $I=100\,$\textmu A (red circles). The red line displays the $T^{3/2}$ behavior.}
	\label{fig:resistivity_comparison}
\end{figure}

In Fig.~\ref{fig:resistivity_comparison}, we show the temperature dependence of the electrical resistivity as determined from pulsed measurements with pulse length $\tau=5\,$\textmu s for the thinner sample.
$R$ is only weakly temperature dependent.
Comparison to the conventional, low-current resistivity measurement indicates an effective sample temperature of $T_\mathrm{eff}\approx5.3\,$K of the CaRuO$_3$ thin film.
The sample temperature increases due to the pulse current.
For the pulsed measurement the measuring current was $I=58\,$mA, while for the conventional, low-current resistivity measurement the measuring current was $I=100\,$\textmu A. Although the measuring current for the pulsed measurement was larger than for the conventional one, it is still much smaller than for the universal region needed.

\subsection{j-E-curve}
\begin{figure}[h!]
	\centering
	\includegraphics[width=0.6\textwidth]{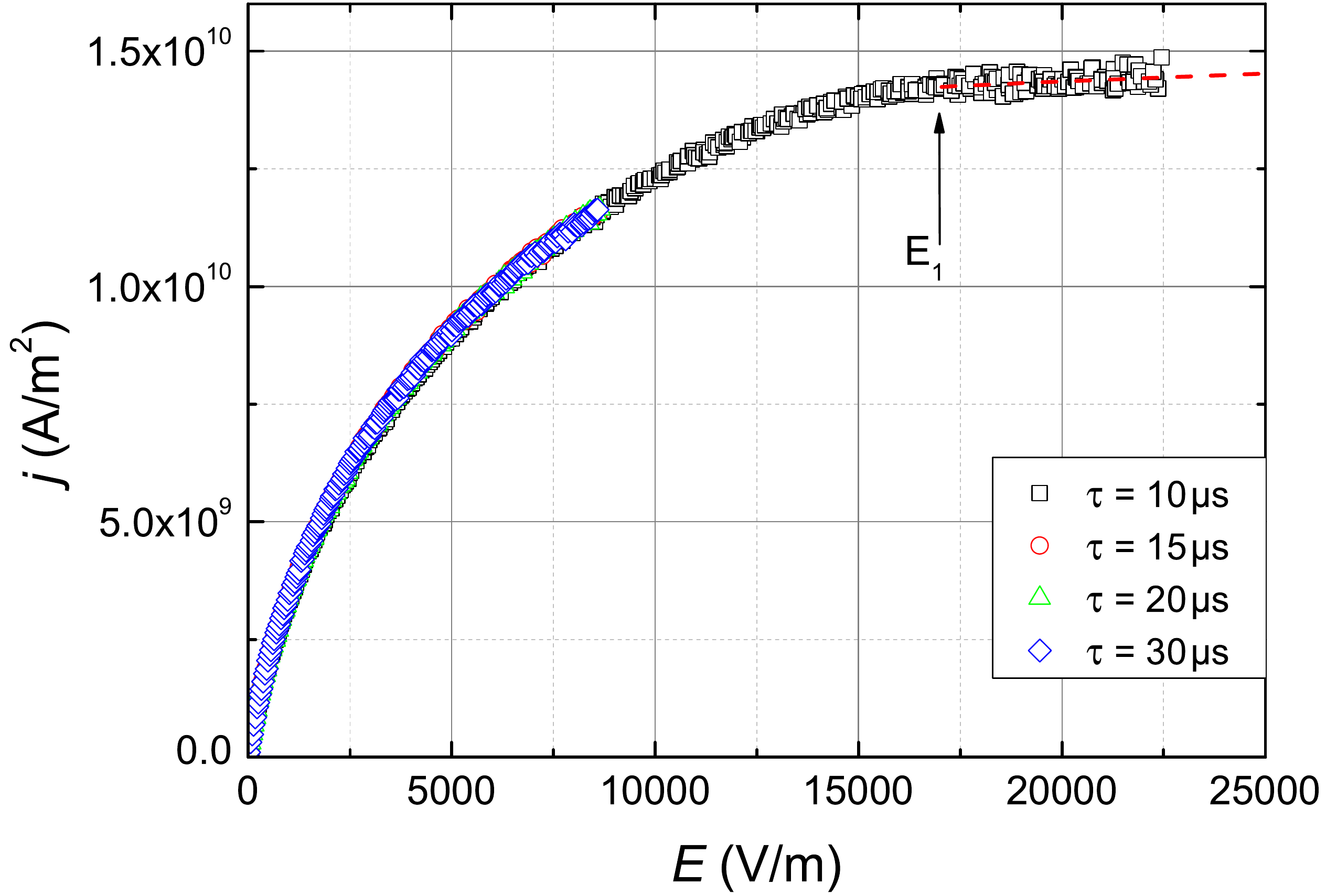}
	\caption{$j$-$E$-curve for CaRuO$_3$ thin film with RRR~$=57$ at 4.2\,K for different pulse lengths.
	The arrow indicates the critical field $E_1$, beyond which universal non-linear behaviour would be expected.
	The red line displays a fit according to $j\propto E^{0.05}$.}
	\label{fig:E_j_curve}
\end{figure}

Figure \ref{fig:E_j_curve} displays the measured $j$-$E$-curve of the thin film with RRR~$=57$ at $4.2\,$K using pulse lengths between $10\,$\textmu s and $30\,$\textmu s.
We observe a strong nonlinear behavior in the shape of the curve for all used pulse lengths.
Furthermore, the curve for $\tau=10\,$\textmu s coincides with all curves with larger pulse lengths.\\
Taking $z=3$ and $\alpha=1.5$, the above equations from Hogan and Green predict a universal $j\propto E^{0.3}$ behavior for fields above a critical field of $E_1=17040\,$V/m.
In contrast to this prediction we observe a very flat curve beyond $E_1$ with exponent close to zero, i.e. a much stronger nonlinearity in the conductivity than expected, which most likely is related to the Joule heating.

\section{Discussion}\label{sec:discussion}
In order to experimentally investigate possible universal nonlinear conductivity behavior for a quantum critical metal, we have performed short-pulse current measurements on high-quality thin films of CaRuO$_3$ on NdGaO$_3$ substrate immersed in liquid helium at low temperatures.
Already at fields much below the threshold value $E_1$ for expected universal nonlinearity, we observe pronounces nonlinear behavior in $j$-$E$-curves, independent of the pulse lengths for pulses as short as 5~\textmu s.
Furthermore, the observed non-linearity at $E>E_1$ is much stronger than predicted for the universal case.
The pulse length dependence of the resistivity shows that joule heating is not yet reduced as the pulse duration~$\tau$ is lowered towards 5~\textmu s.
We thus conclude that in our experiment, we can only observe the high-$\tau$ constant temperature regime indicated in Fig.~\ref{fig:resistivity_schematic}c and Joule heating is responsible for the vast bulk of the observed nonlinear conductivity.

The assumption of the substrate acting as a perfect heat sink seems may be too simplistic in this case.
Despite epitaxial growth and high substrate heat capacity, a finite thermal boundary resistance between film and substrate and the finite thermal conductivity of the substrate may hinder an efficient transport of the heat dissipated in the film.

Our results show that it is experimentally challenging to verify the predicted nonlinear conductivity.
Pulses much shorter than $5\,$\textmu s are needed.
However, in order to apply the formalism for steady-state non-equilibrium, the pulses must not get shorter than intrinsic time scales of the (slow) fluctuations within the quantum critical regime.
Recent THz spectroscopy indicates that the frequency dependent real part of the resistivity approaches the dc resistivity below about  0.2\,THz which corresponds to 5\,ps~\cite{schneider2013low}.
Thus, short-pulse experiments on ns time scales may be able to approach the universal nonlinear regime.

\section*{Appendix}
\begin{figure}[h!]
	\centering
	\includegraphics[width=0.7\textwidth]{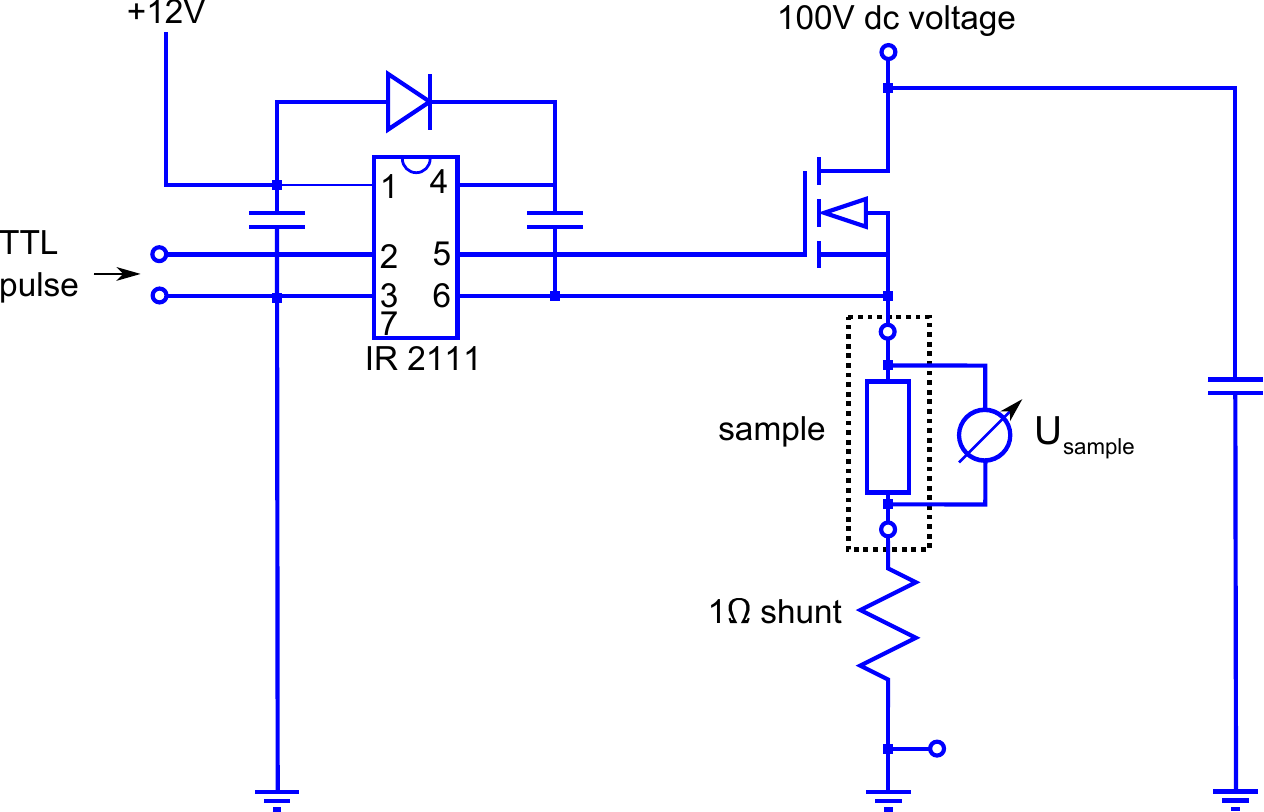}
	\caption{Detailed version of the used circuit diagram.
	The ports of the IR2111 are numbered as: 1 V$_\mathrm{CC}$, 2 IN, 3 COM, 4 V$_\mathrm{B}$, 5 HO, 6 V$_\mathrm{S}$, 7 LO.}
	\label{fig:detailed_circuit}
\end{figure}

Figure \ref{fig:detailed_circuit} displays a more detailed version of the circuit used for producing short current pulses.
By using an IR2111 gate driver to level-shift the pulse from the signal generator, the transistor can be located at the high side of the sample, which avoids large voltage swings at the measurement outputs to the ADC.

\end{document}